\documentclass[%
 reprint,
 amsmath,amssymb,
 aps,
]{revtex4-2}

\usepackage{graphicx}
\usepackage{dcolumn}
\usepackage{bm}

\begin{document}

\preprint{APS/123-QED}

\title{Conceptual Design of a Low-Energy Ion Beam Storage Ring and\\ a Recoil Separator to Study Radiative Neutron Capture by Radioactive Ions}

\author{Kihong Pak}
\affiliation{Department of Nuclear Engineering, Hanyang University,\\ Seoul 04763, Republic of Korea}

\author{Barry Davids}
\email{davids@triumf.ca}
\affiliation{
 TRIUMF, 4004 Wesbrook Mall, Vancouver, BC V6T 2A3, Canada
}%
\affiliation{
 Department of Physics, Simon Fraser University, 8888 University Drive, Burnaby, BC V5A 1S6, Canada
}%
\author{Yong Kyun Kim}
\email{ykkim4@hanyang.ac.kr}
\affiliation{%
 Department of Nuclear Engineering, Hanyang University,\\ Seoul 04763, Republic of Korea
}%

\begin{abstract}
Recently, the TRIUMF Storage Ring (TRISR), a storage ring for the existing Isotope Separator and Accelerator-I (ISAC-I) radioactive ion beam facility at TRIUMF, was proposed. It may be possible to directly measure neutron-induced radiative capture reactions in inverse kinematics by combining the ring with a high-flux neutron generator as the neutron target. Herein, we present the conceptual design of a low-energy ion storage ring as well as a fusion product extraction system with a Wien filter and recoil separator for detecting neutron capture products based on ion optical calculations and particle-tracking simulations.
\end{abstract}

\maketitle


\section{\label{sec:intro}INTRODUCTION}

The origins and abundances of the elements have been quantitatively studied for more than six decades \cite{B2FH1957,cameron57}. It is well known that hydrogen, helium, and trace quantities of lithium were produced in primordial nucleosynthesis shortly after the Big Bang \cite{pitrou18,fields20}. Heavier elements were produced from the lightest nuclei through nucleosynthetic processes such as fusion reactions in stars.

Nuclei heavier than iron ($Z=26$) are primarily synthesized by photodisintegration reactions in the $p$ process and/or by radiative neutron capture reactions $(n,\gamma)$ occurring in the slow ($s$), rapid ($r$), or intermediate ($i$) neutron capture processes \cite{rauscher20}.
Radiative neutron capture reactions are those in which an atomic nucleus absorbs a neutron to form a heavier nucleus with the emission of one or more $\gamma$-rays. The classical $s$ process begins with $^{56}$Fe as a seed nucleus and climbs along the valley of $\beta$ stability up to $^{209}$Bi \cite{kaeppeler11}. The $r$ process drives nuclei far to the neutron-rich side of the stability line \cite{cowan21}. The path of the more recently introduced $i$ process lies in between, as its neutron density lies between those of the $s$ and $r$ processes \cite{cowan77,denissenkov17}.

For several decades, the neutron capture cross-sections of $s$ process nuclides have been measured using time-of-flight \cite{Beer1997} and activation techniques \cite{Kappeler1990, Kappeler1996}. Most of the $r$ process involves nuclides far from the valley of stability, so the products are too short-lived to be directly measured. Indirect techniques such as the surrogate reaction method have been applied \cite{Boyer2006, escher12}. Therefore, it is necessary to obtain as much experimental data as possible away from the stability line to assist in theoretical extrapolations of nuclear properties. Generally, the shorter the half-life of the nuclei under investigation, the more difficult it is to prepare a radioactive sample and perform measurements.

In the case of proton- and $\alpha$-induced capture reactions, inverse kinematics can be used; however, for neutron-induced reactions, this approach requires a (radioactive) neutron target. Reifarth \emph{et al.} \cite{Reifarth14, Reifarth17} suggested combining an ion storage ring with a research reactor as a neutron source to directly measure neutron-induced reactions in inverse kinematics. With a thermal neutron density of $2\times10^{10}$~cm$^{-2}$ from the reactor and $10^{13}$ ions s$^{-1}$ passing through a 500~cm long interaction region ($10^7$ stored ions circulating in a ring with a frequency of 1 MHz), the achievable luminosity of $2\times 10^{23}$~cm$^{-2}$~s$^{-1}$ would be large enough to measure reactions with cross-sections as small as a few mb within a day. Because the neutron capture reaction product has the same momentum as the primary beam of mass number $A$ but its speed is only $A/(A+1)$ times as large, it is possible that Schottky spectroscopy could be used to detect the synthesized ions through the change in the revolution frequency.

Recently, a storage ring coupled with the existing ISAC facility, combined with a high-flux neutron generator, was proposed at TRIUMF based on this idea \cite{dillmann23}. This new low-energy storage ring, the TRISR, would be able to utilize high-intensity radioactive ion beams ($\geq10^8$ s$^{-1}$) with an energy range of 0.15~$A$~MeV up to 1.8~$A$~MeV for $A/q \leq 7$. The neutron generator would need to provide a moderated flux of thermal neutrons with an areal number density exceeding $10^8$~cm$^{-2}$ in a compact geometry. 

A new method was proposed to extract the reaction products from the ring using a velocity (Wien) filter and a magnetic septum and detect them using a recoil separator. The maximum electric field strength of such a Wien filter was assumed to be 35~kV~cm$^{-1}$, on the basis of extensive experience operating Wien filters at TRIUMF \cite{beveridge85}. In this system, ion beams stored in the ring rotate and pass through the neutron reaction target at frequencies of a few hundred kHz, sometimes producing neutron capture reaction products. The recoils would be removed from the ring and transported to the focal plane of the recoil separator, arriving within $\mu$ s after the reaction occurred, to be identified and counted. This paper presents conceptual designs of a low-energy storage ring and recoil separator for the measurement of neutron capture cross-sections. Ion optical calculations were performed to determine the locations and specifications of the electromagnetic components of the ring and recoil separator. Recoil trajectories after the Wien filter were evaluated using particle-tracking simulations to determine the location and specifications of a septum magnet to transfer recoils to the recoil separator.

\section{Conceptual Design of \\ a Storage Ring}
\begin{figure}[ht]
 \includegraphics[width=0.5\textwidth]{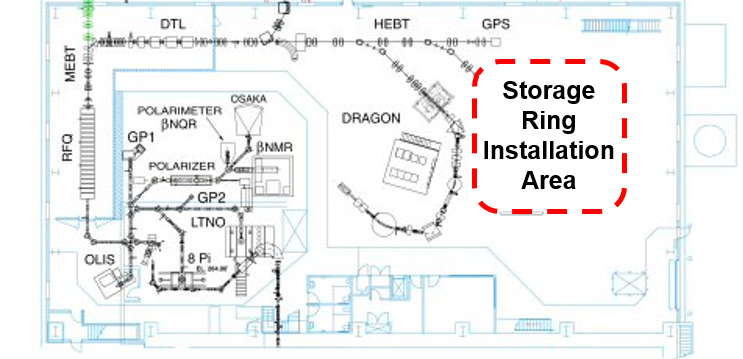}
\caption{Plan view of the ISAC-I experimental hall at TRIUMF. A possible location of a storage ring that would be coupled to the existing ISAC radioactive beam facility is shown.} \label{fig:isac}
 \end{figure}

Because the TRISR would likely be installed within the existing space in the ISAC-I experimental hall depicted in Figure \ref{fig:isac}, the circumference of the ring cannot exceed 50 m. The size and essential devices required in the lattice are similar to those of the former Test Storage Ring (TSR) that was operated at the Max Planck Institute in Heidelberg \cite{Grieser2012}. However, while the TSR had a symmetric square design, the TRISR is rectangular in shape because a drift length is required after the integral Wien filter to increase the separation of the recoils from the primary beam. The storage ring has an injection system and a Wien filter, followed by a magnetic septum in a long straight section; an electron cooler is under consideration. A neutron generator would be placed in a short section downstream from the injection system, forming the reaction area. In the other short section, Schottky pickups or other devices may be placed according to scientific and technical requirements. The designs of these components are beyond the scope of this work.

\subsection{Ion Optical Calculations}
\begin{figure}[h] \centering
 \includegraphics[width=0.5\textwidth]{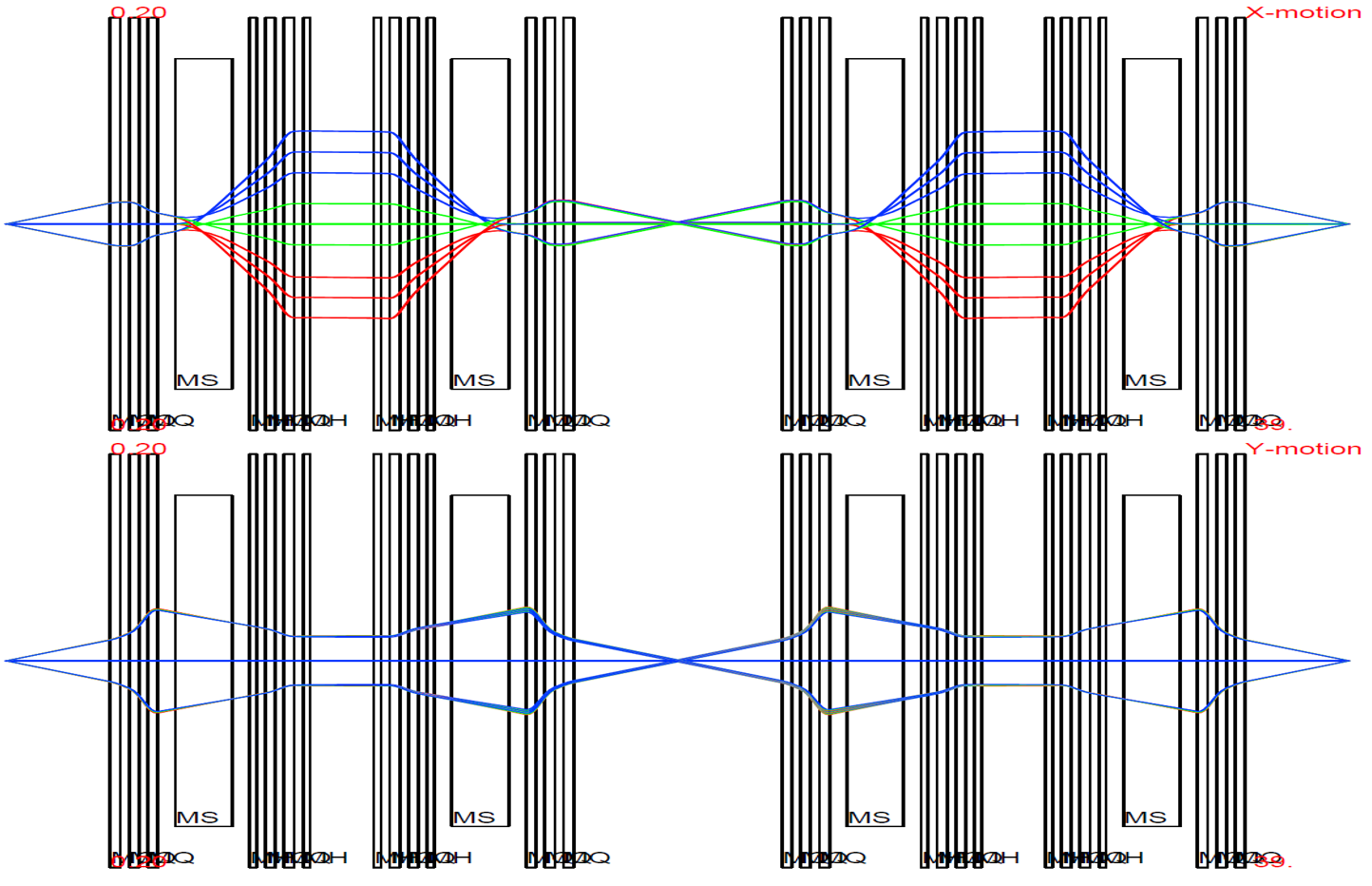}
\caption{The results of ion optical calculations for a single turn in the ring. The initial beam size and angular divergences are $x,y$: $\pm1$~mm and $x',y'$: $\pm2.8$~mrad, respectively. The upper and lower plots show the trajectories in the horizontal and vertical directions, respectively, with momentum deviations of +2\% (blue), 0\% (green) and -2\% (red). MQ: quadrupole, MH: sextupole, MS: dipole.} \label{fig:cosy_noff}
 \end{figure}

To determine the locations and specifications of the electromagnets, such as dipoles, quadrupoles, and sextupoles, to store the ion beam in the ring, ion optical calculations were performed using COSY INFINITY 10.0 \cite{makino05,cosy_ref}. The transfer matrix elements were calculated up to the fifth order and fringe field effects were included. Fitting was performed by varying the locations, field strengths, and combinations of magnets to achieve $(x,x)=(x',x')=(y,y)=(y',y')=1$ and $(x,x')=(y,y')=(x,\delta_p)=0$ and minimize the corresponding higher-order terms at the middle and end of the ring. Here, $x$, $y$, $x'$, $y'$, and $\delta_p$ are the horizontal and vertical positions, horizontal and vertical angles, and relative momentum deviation, respectively. Figure \ref{fig:cosy_noff} shows the results of the ion optical calculations for the storage ring. $^{80}$Zn$^{15+} (Z=30, N=50)$ at 0.15 MeV/nucleon with a normalized (4 root mean square) transverse emittance of $0.2 \pi$~$\mu$m \cite{laxdal14} was arbitrarily selected as the circulating beam for the calculations.

\begin{figure}[htb!] \centering
 \includegraphics[width=0.5\textwidth]{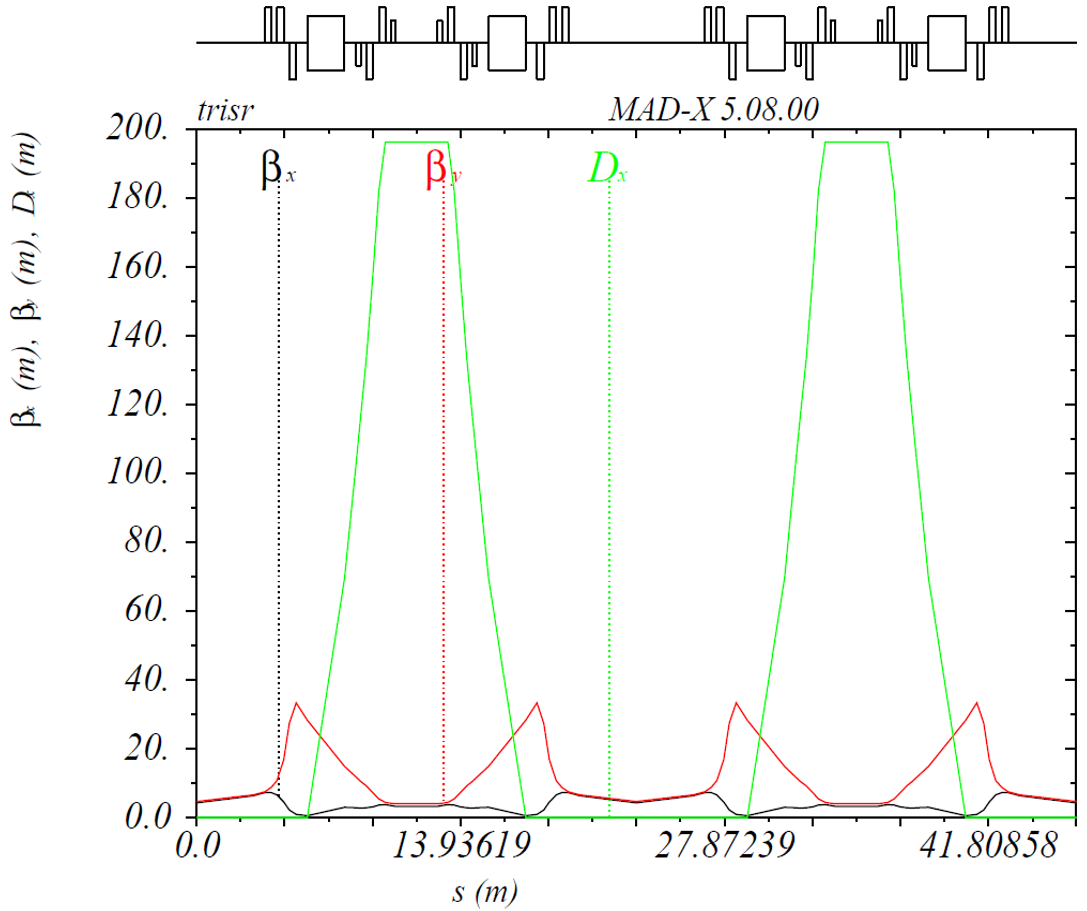}
\caption{Beta functions in the horizontal (black) and vertical (red) directions and dispersion function in the horizontal direction (green), which depend on the path length along the optic axis $s$.} \label{fig:madx_plot}
 \end{figure}

The beam starts in the middle of the first straight section and ends at the same point after a single rotation with the same angle and no momentum dispersion. The reaction area was positioned with a 1~m long neutron target in the second straight section and a Wien filter in the third section. Thus, the neutron capture products created in the second section can be delivered to the next section along the same trajectory as that of the primary beam ions, as they have nearly identical magnetic rigidities $B\rho$. The recoils can then be separated from the un-reacted beam and deflected by the Wien filter. The recoils are delivered to the recoil separator through a magnetic septum, whereas the primary beam ions are stored in the ring and circulated until they react with neutrons, decay, or are lost via charge-changing collisions. The ring consists of four 90$^{\circ}$ dipoles, 20 quadrupoles, and eight sextupoles; it exhibits a mirror-symmetric structure to reduce higher-order aberrations \cite{Erdelyi2007}. Its field-free straight sections are 7.2~m and 1.8~m long, and the total length of the ring is 46.5~m.
To verify the properties of the system described above, we used the Methodical Accelerator Design-X program \cite{madx} to calculate the ion optical parameters, such as $\beta_{x,y}$, $\eta_{x,y}$, and $\nu_{x,y}$. Figure \ref{fig:madx_plot} shows plots of the beta and dispersion functions in the ring, and Table \ref{table:parameter} presents the parameters of the designed ring. The Courant-Snyder parameters are the same at the beginning and end of the ring.

\begin{table}[htb!]
\caption{\label{table:parameter}%
Ion Optical Parameters of the Ion Storage Ring}
\begin{ruledtabular}
\begin{tabular}{lcdr}
\textrm{Parameter}&
\textrm{Value} \\
\colrule
Circumference (m)   & 46.5  \\
$\beta_{max,x}$ (m) & 7.3   \\
$\beta_{max,y}$ (m) & 33.3  \\
$\eta_{max,x}$ (m)  & 196.4 \\
\end{tabular}
\end{ruledtabular}
\end{table}

\subsection{Evaluation of Recoil Trajectories after the Extraction System}
A Wien filter separates particles whose speeds differ from that of the selected particle, $v_o=E/B c$, where $E$ and $B$ are the strengths of its orthogonal electric and magnetic fields, respectively. A particle with speed $v_o$ passes through the filter without deflection because the Lorentz force induced by the $E$ and $B$ fields vanishes; however, a particle with speed $v\neq v_o$ is deflected by the Lorentz force, $\mathbf{F}=q(\mathbf{E}+ \frac{\mathbf{v}}{c}\times \mathbf{B})$. To separate the recoils from the beam, a Wien filter with an effective length of 2~m is placed downstream of a quadrupole triplet at the beginning of the third straight section. After neutron capture, but prior to the emission of one or more photons in the radiative capture reaction, the beam and recoil ions have approximately the same momentum but different masses and speeds, as indicated in Equation \ref{eqn1}.

\begin{equation} \label{eqn1}
    p_0=p;\ \frac{Av_o}{q_o}=\frac{(A+1)v}{q_o};\ v=\frac{A}{(A+1)}v_o,
\end{equation}
where $p_o$ and $p$ are the momenta of the beam and recoil, respectively; $A$ is the mass number of the beam; $v_o$ and $v$ are the speeds of the beam and recoil, respectively; and $q_o$ is the charge of the beam.

For a circulating beam of $^{80}$Zn at 0.15~MeV/nucleon ($v_o \approx 0.018c$), the $^{81}$Zn recoil has a speed of $v\approx 0.9877 v_o$ after a neutron is captured. The deflected recoil should be incident on a magnetic septum located downstream of the Wien filter and delivered to the recoil separator. Therefore, the trajectories of the beam and recoil after passing through the Wien filter were evaluated and used to determine the location of the magnetic septum. The software package G4beamline \cite{G4bl_manual}, which is based on the GEANT4 toolkit \cite{geant}, was used to track the individual beam and recoil ion trajectories through the storage ring and Wien filter.

\begin{figure}[htb!] \centering
 \includegraphics[width=0.5\textwidth]{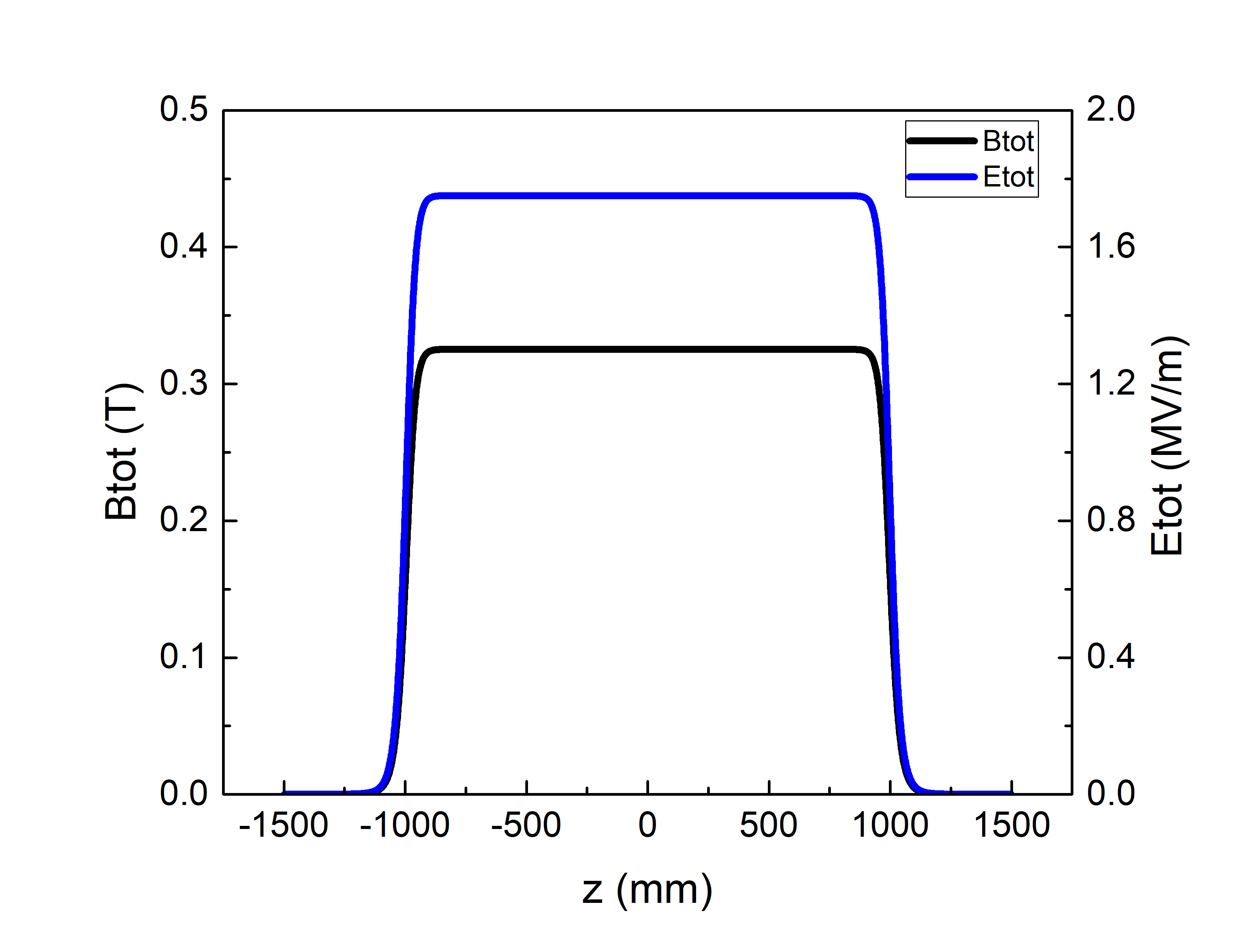}
\caption{The orthogonal electric and magnetic fields of the Wien filter, including the fringing fields.} \label{fig:WFfringefield}
 \end{figure}

The Wien filter was implemented at the beginning of the third straight section with a length of 2~m and an aperture of $100 \times 100$~mm$^2$. Orthogonal electric and magnetic fields were applied throughout the gap using a three-dimensional field map. The fringing fields were parameterized using Enge functions with coefficients ($a_1, a_2, ..., a_6$) derived from the measured magnetic field of a device recently manufactured by Danfysik. The effective length of the filter was evaluated as 2000.7 mm. The coefficients are listed in Table \ref{table:WFcoefficients}, and the electric and magnetic field profiles are shown in Figure \ref{fig:WFfringefield}.

\begin{table}[htb!]
\caption{\label{table:WFcoefficients}%
Enge coefficients of the Wien filter fringing fields.}
\begin{ruledtabular}
\begin{tabular}{cdr}
\textrm{Enge coefficient}&
\textrm{Value} \\
\colrule
$a_1$ &  -9.245                 \\
$a_2$ &  -0.610                 \\
$a_3$ &  0.110                  \\
$a_4$ &  -0.003                 \\
$a_5$ &  -3.62 \times 10^{-5}   \\
$a_6$ &  3.4 \times 10^{-6}     \\
\end{tabular}
\end{ruledtabular}
\end{table}

The position distributions of the recoils at the reaction target are the same as those of the beam; however, the angular divergences are slightly larger, owing to the emission of one or more photons. We assume that the capture reaction is followed by the emission of a single photon, thereby maximizing the associated momentum transfer. The maximum recoil angle $\theta$ and spread $\Delta p / p$ about the central recoil momentum $p$ can be calculated using Equation \ref{eqn2}, where $Q$ is the reaction $Q$ value; $E$ is the relative kinetic energy; and $m_b$, $m_t$, and $m_r$ are the masses of the beam, target, and recoils, respectively \cite{BruneDavids15}.

\begin{eqnarray} \label{eqn2}
\frac{\Delta p}{p} = \tan \theta_{max}= \frac{m_rc^2(\sqrt{1+\frac{2(Q+E)}{m_rc^2}}-1)}{\sqrt{2m_bc^2 (\frac{m_b+m_t}{m_t})E}} \nonumber \\
\approx \frac{Q+E}{\sqrt{2m_bc^2 (\frac{m_b+m_t}{m_t})E}}
\end{eqnarray}

\begin{figure}[htb!] \centering
 \includegraphics[width=0.5\textwidth]{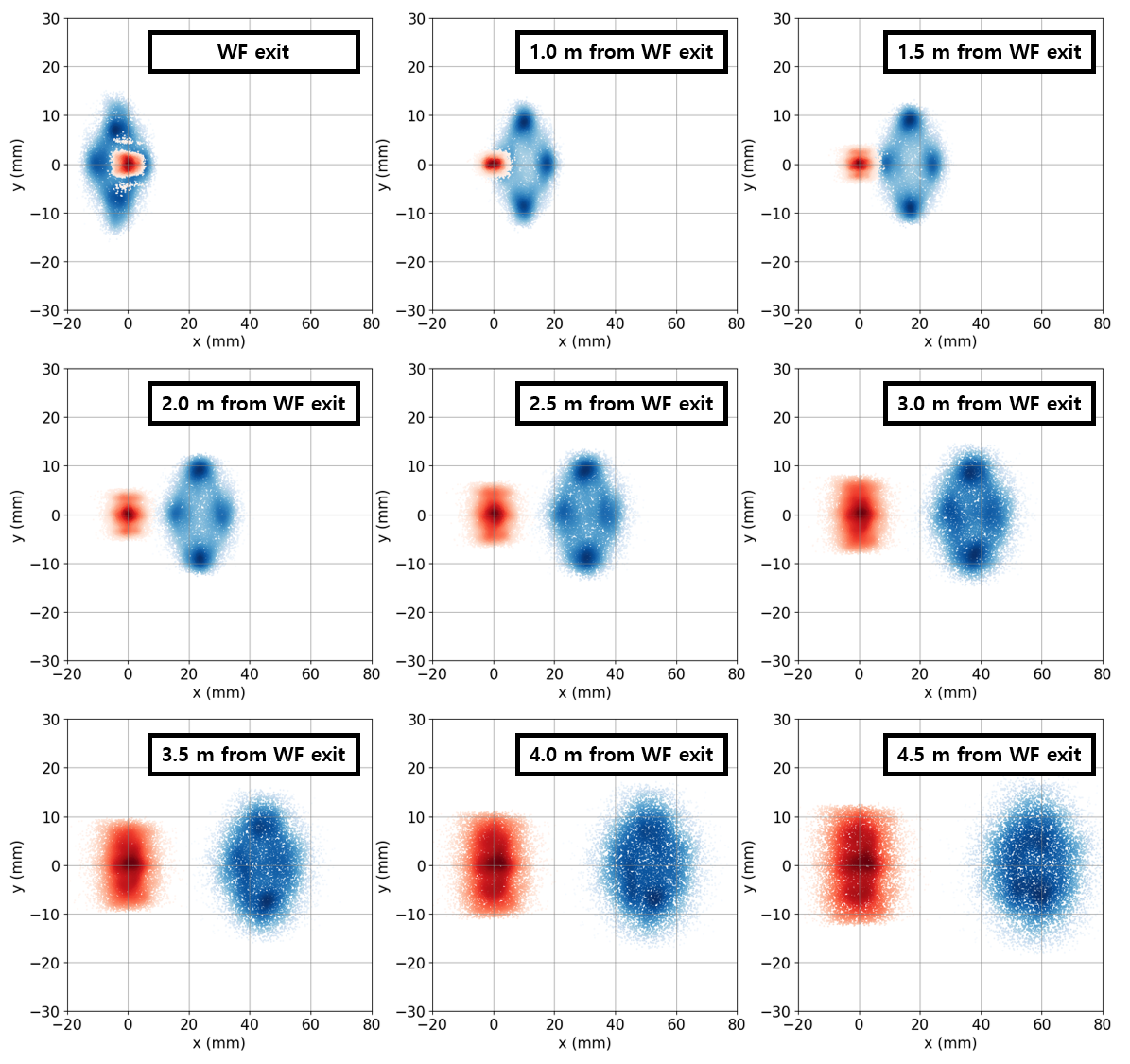}
\caption{Position distributions of the beam (red) and recoil (blue) ions after passing through the Wien filter at various downstream distances.} \label{fig:position}
 \end{figure}

The maximum scattering angle for the case of the $^{81}$Zn recoils we considered is 2.1~mrad, corresponding to a momentum transfer of 2.77~MeV/c. The trajectories of the beam and recoil ions exiting the Wien filter were calculated to determine the location of the magnetic septum. Figure \ref{fig:position} shows the position distributions of the beam and recoils at different distances from the Wien filter exit. After a drift length of approximately 4~m, the products are dispersed by about 40 mm from the beam in the horizontal direction and can be extracted using a magnetic septum, in which a separation of 2-20~mm between the homogeneous and field-free regions is typical \cite{Barnes2011}.

\begin{figure}[htb!] \centering
 \includegraphics[width=0.5\textwidth]{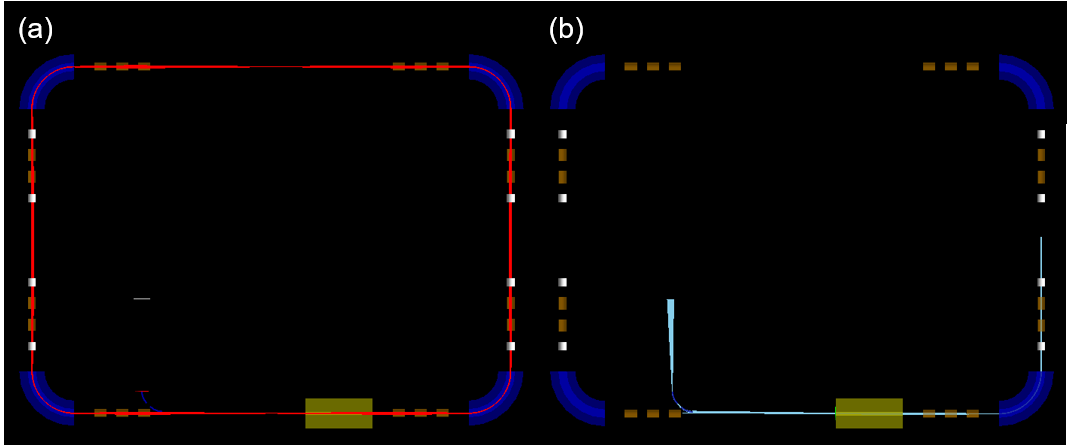}
\caption{(a) The ion beam trajectories throughout the storage ring. The beam starts at the middle of the first (top) straight section with $x,y=\pm1$~mm, $x',y'=\pm2.8$~mrad, and $\delta_ p=\pm2\%$. (b) The recoil trajectories from the reaction target to the extraction system. The recoils start at the middle of the second (right) straight section with $x,y=\pm1$~mm, $x',y'=\pm2.8$~mrad, and $\delta_p=\pm2\%$. 10,000 particles were sampled from each distribution.} \label{fig:g4bl}
 \end{figure}

\begin{figure}[htb!] \centering
 \includegraphics[width=0.5\textwidth]{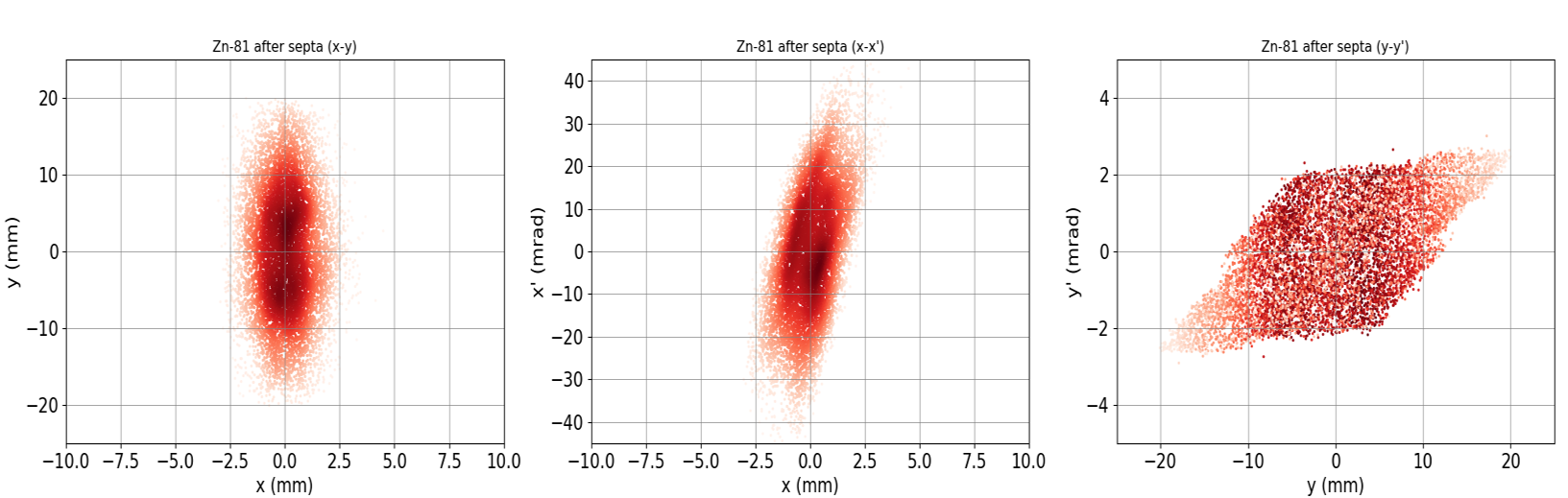}
\caption{$x-y$ (left), $x-x'$ (middle), and $y-y'$ (right) phase space distributions of the recoils after they are extracted by the magnetic septum and emerge from the following dipole magnet.} \label{fig:phasespace}
 \end{figure}

The position of the magnetic septum was chosen to be 4.3 m downstream of the Wien filter, considering the dispersion and potential interference with the quadrupole. The extraction system consists of a 45$^\circ$ magnetic septum, which has a small aperture and a typical separation between the uniform and vanishing field regions, followed by a 45$^\circ$ dipole magnet to deliver the reaction products to the recoil separator. Figure \ref{fig:g4bl} shows the trajectories of the beam ions and recoils simulated using the G4beamline program. The beam is stored in the ring, continuously circulating while the recoil is separated from it by the Wien filter and extracted through the subsequent septum. Beam ions with momenta that differ slightly from the reference momentum are deflected by the Lorentz force in the Wien filter. However, an electron cooler situated in the first straight section could compensate for these momentum deviations, ensuring that the beam continues to circulate stably. The phase space distributions of the recoils after the extraction system are presented in Figure \ref{fig:phasespace} and were used as an input for designing the recoil separator. The properties of the electromagnets in the ring are shown in Table \ref{table:ringmagnetproperties}.

\begin{table*}
\caption{\label{table:ringmagnetproperties}The properties of the electromagnets in the storage ring. $B_{max}$ indicates the magnitude of the peak magnetic field or gradient, $L_{eff}$ is the effective length, and $\rho$ is the radius of curvature. Here $n$ is 0 for the dipole and septum magnets and the Wien filter, 1 for the quadrupoles, and 2 for the sextupoles.}
\begin{ruledtabular}
\begin{tabular}{ccccccc}
Element & $B_{max}$ & Aperture & $L_{eff}$ & Bending & $\rho$ & Quantity \\
 & (T/m$^n$) & (mm) & (mm) & Angle ($^{\circ}$) & (mm) \\
\hline
Dipole      & 0.24 & 250$\times$80 & 1963 & 90 & 1250 & 4 \\
Quadrupole  & 0.68 & 200 & 350 &&&20\\
Sextupole   & 0.50 & 200 & 250 &&&8\\
Wien Filter & 0.65 & 100$\times$100 & 2000 &&&1\\
Septum & 0.59 & 60$\times$100 & 393 & 45 & 500 &1\\
Dipole & 0.59 & 60$\times$100 & 393 & 45 & 500 &1\\
\end{tabular}
\end{ruledtabular}
\end{table*}

\section{Conceptual design of a recoil separator}
In their paper on measuring neutron-induced reactions in inverse kinematics, Reifarth and Litvinov suggested using Schottky spectroscopy to detect the appearance of synthesized ions that remain in the storage ring, as well as particle detectors located after a dipole magnet to detect those that do not \cite{Reifarth14}. However, the inclusion of a Wien filter in the ring, followed by a magnetic septum and recoil separator, would offer an alternative method for detecting fusion products. The aim would be to continuously extract recoil ions from the circulating stored beam at a point close to the neutron target. Thus, reaction products would be removed from the ring within a very short time after the reaction occurred, on the order of $\mu s$, and transported to a detection area at the focal plane of the recoil separator, where standard particle identification techniques would be applied. The advantage of this approach is that the detection efficiency does not depend on the recoil lifetime, which can be very short. In addition, negligible energy loss and scattering would occur for the neutrons in the target, leading to minimal amounts of scattered or leaky beam that could pass through the recoil separator. \\

The design criteria and configuration of the recoil separator were based on those of the DRAGON and EMMA facilities at TRIUMF \cite{Hutcheon2003,Davids2019}. DRAGON is used to measure proton and $\alpha$ radiative capture reactions by electromagnetically separating the reaction products from a beam that impinges on a windowless gas target. The beam and other products with masses and charge states that differ from those of the selected recoil are removed by slit systems following a dipole magnet and an electrostatic deflector. EMMA is a vacuum-mode recoil mass spectrometer that separates the recoils of nuclear reactions from the ion beam, focuses their energy and angle, and disperses them in a focal plane according to their mass/charge ($m/q$) ratios. The $m/q$ dispersion of EMMA is variable and it is typically operated at 10 mm/$\%$. \\

\begin{figure}[htb!] \centering
 \includegraphics[width=0.5\textwidth]{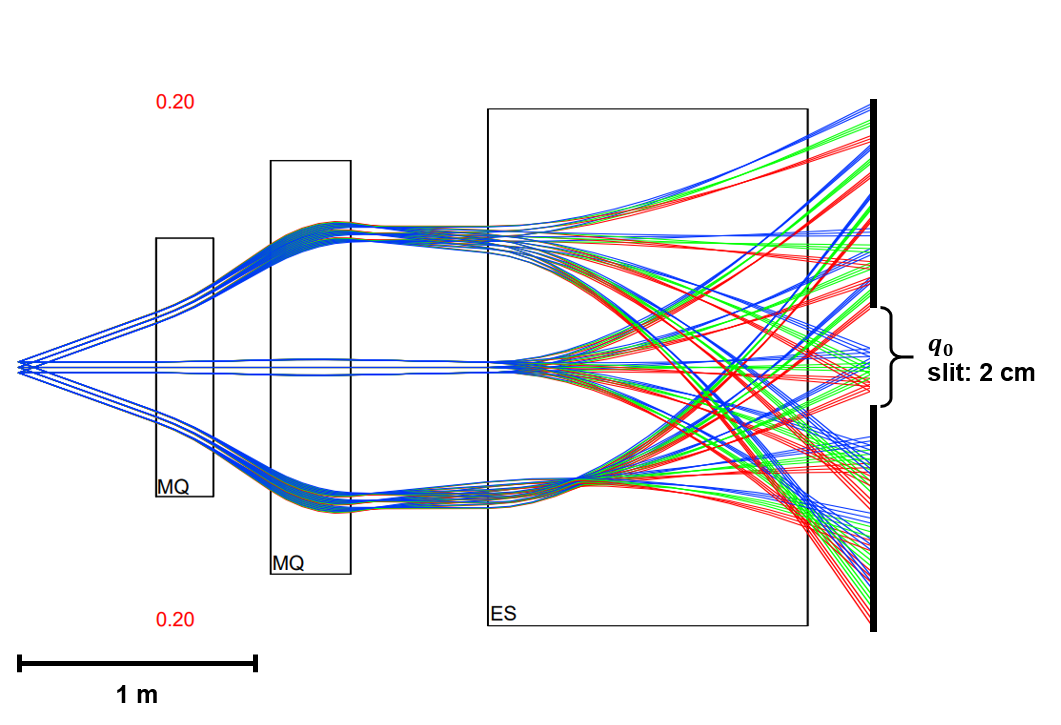}
\caption{The optics of charge state selection in the recoil separator using an electrostatic deflector in the horizontal direction. The beam ions start from a spot with a size of $\pm 2.1\ mm$ and angular divergences of $\pm 31.9$ mrad in both transverse directions, a momentum deviation $\Delta p/p$ of $\pm 2\%$, and charge states of $q_0-2$, $q_0-1$, $q_0$, $q_0+1$, and $q_0+2$. MQ denotes a quadrupole magnet and ES an electrostatic deflector.} \label{fig:recoilSeparatorChargeSelection}
 \end{figure}

\begin{figure}[htb!] \centering
 \includegraphics[width=0.5\textwidth]{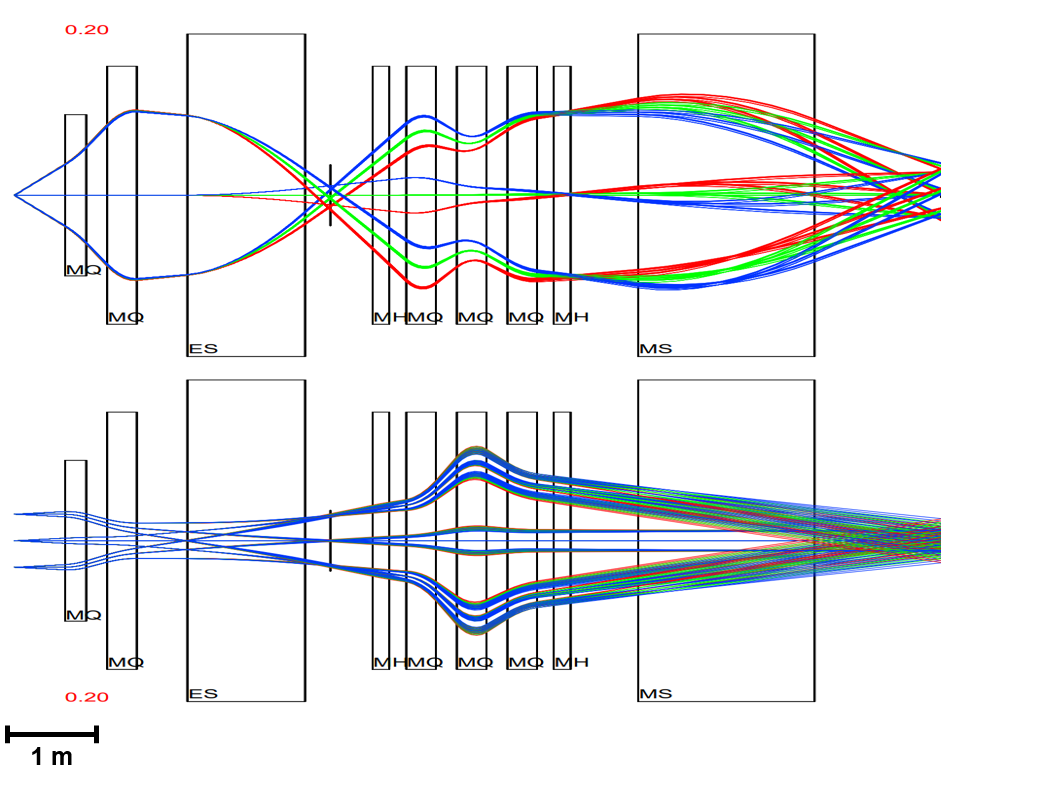}
\caption{The optics of the recoil separator in the horizontal (top) and vertical (bottom) directions. The beam starts 1~m before the first quadrupole magnet with $x= \pm 2.1$~mm, $y= \pm 16.5$~mm, $x'= \pm 31.9$~mrad, $y'= \pm 2.7$~mrad, $\Delta p/p= \pm2\%$, and $\Delta m/m= \pm1.2\%$. MQ denotes a quadrupole magnet, MH a sextupole magnet, ES an electrostatic deflector, and MS a dipole magnet.} \label{fig:recoilSeparator}
 \end{figure}

The TRISR recoil separator presented here was conceptually designed to stop any ions with charge states that differ from the selected charge state of the recoils, $q_0$, using a slit system located after an electrostatic deflector, and to disperse recoils with charge state $q_0$ according to the mass at the focal plane downstream of a dipole magnet. The separator consists of five quadrupoles, two sextupoles, a 40$^\circ$ electrostatic deflector, and a 60$^\circ$ dipole magnet. The deflection angles of the dipoles were chosen to achieve an $m/q$ dispersion similar to that of EMMA. The total length of the recoil separator beam line is 11.0~m. Figure \ref{fig:recoilSeparatorChargeSelection} shows the trajectories of recoils with five different charge states, $q_0-2$, $q_0-1$, $q_0$, $q_0+1$, and $q_0+2$. Figure \ref{fig:recoilSeparator} shows the optics of the recoil separator and Table \ref{table:seperatormagnetproperties} lists the properties of the separator electromagnets. 

\begin{figure}[htb!] \centering
 \includegraphics[width=0.5\textwidth]{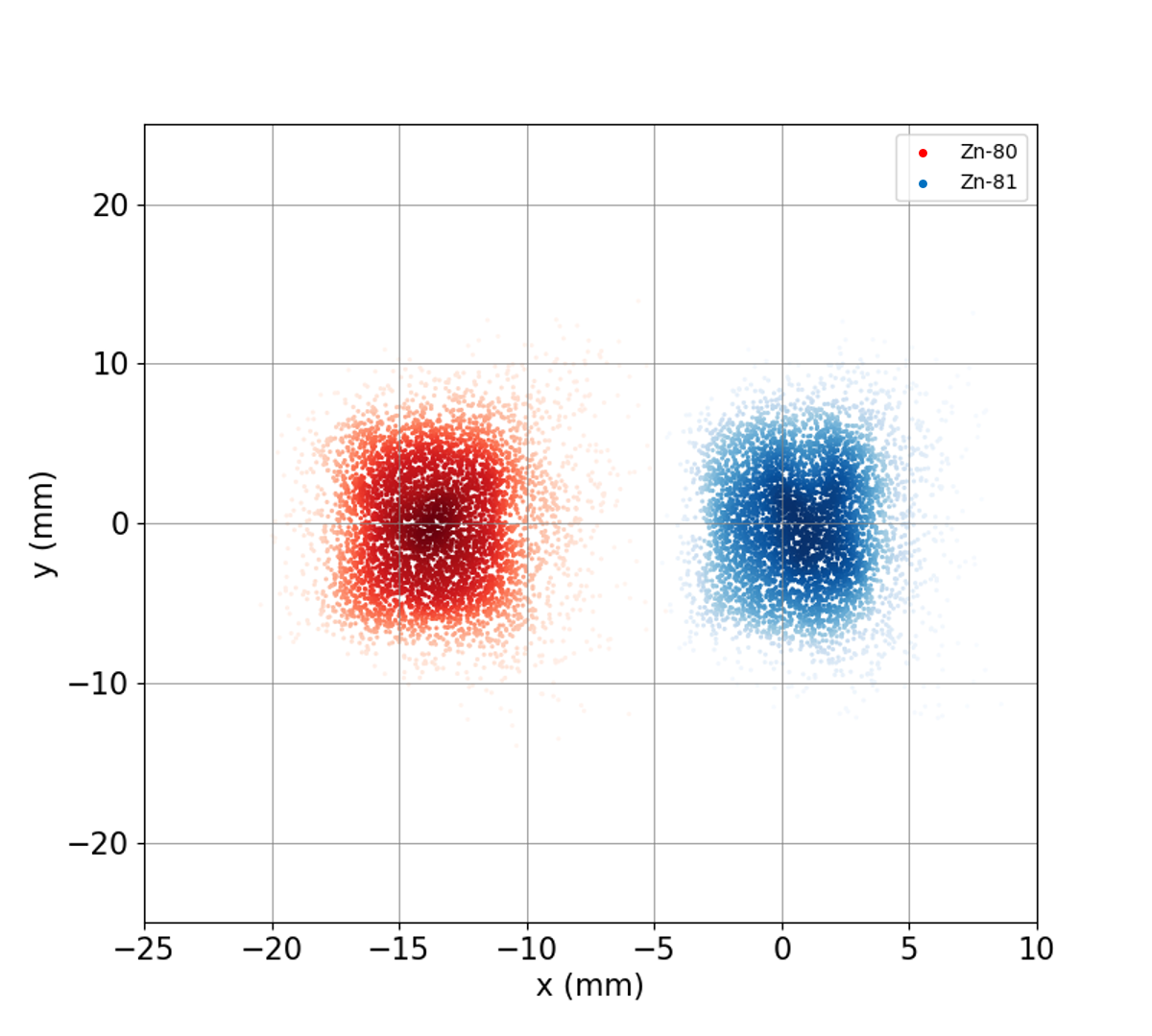}
\caption{Calculated position distributions of beam and recoil ions in the focal plane of the recoil separator.
} \label{fig:beamspot}
 \end{figure}

The initial beam envelope of $x=\pm 2.1$~mm, $x'=\pm 31.9$~mrad, $y=\pm16.5$~mm, and $y'=\pm2.7$~mrad was determined from the $\pm2\sigma$ values of the position and angular distributions of the reaction products emerging from the dipole magnet downstream of the magnetic septum. Ions with the selected charge state $q_0$ are dispersed according to the mass by the dipole magnet of the recoil separator and focused in terms of energy and angle in a focal plane with an $m/q$ dispersion of 11.4 mm/$\%$. The calculated positions of the recoils and beam ions in the focal plane are shown in Figure \ref{fig:beamspot}, assuming that some scattered beam ions are injected into the separator.

\begin{table*}[t]
\caption{\label{table:seperatormagnetproperties}The properties of the electromagnets in the recoil separator used in the ion optical calculations. $B_{max}$ and $E_{max}$ indicate the peak strength of the required magnetic field (or gradient) and the maximum electric field, respectively, $L_{eff}$ is the effective length, and $\rho$ is the radius of curvature. Here, $n$ is 0 for the electric and magnetic dipole, 1 for the quadrupoles, and 2 for the sextupoles. Q denotes a quadrupole magnet, ES an electrostatic deflector, and MS a dipole magnet.}
\begin{ruledtabular}
\begin{tabular}{cccccccc}
Element & $B_{max}$ & $E_{max}$ & Aperture & $L_{eff}$ & Deflection & $\rho$ & Quantity \\
& (T/m$^n$) & (MV/m) & (mm) & (mm) & angle ($^{\circ}$) & (m) \\
\hline
Q1        & 1.18  &   & 50  & 250  &    &   & 1 \\
Q2-5      & 0.76  &   & 80  & 350  &    &   & 4 \\
Sextupole & 2.06  &   & 80  & 300  &    &   & 2 \\
ES        &       & 1 & 100 & 1396 & 40 & 2 & 1 \\
MS        & 0.65  &   & 100 & 2094 & 60 & 2 & 1 \\
\end{tabular}
\end{ruledtabular}
\end{table*}

\section{Summary and conclusion}
\begin{figure}[htb!] \centering
 \includegraphics[width=0.5\textwidth]{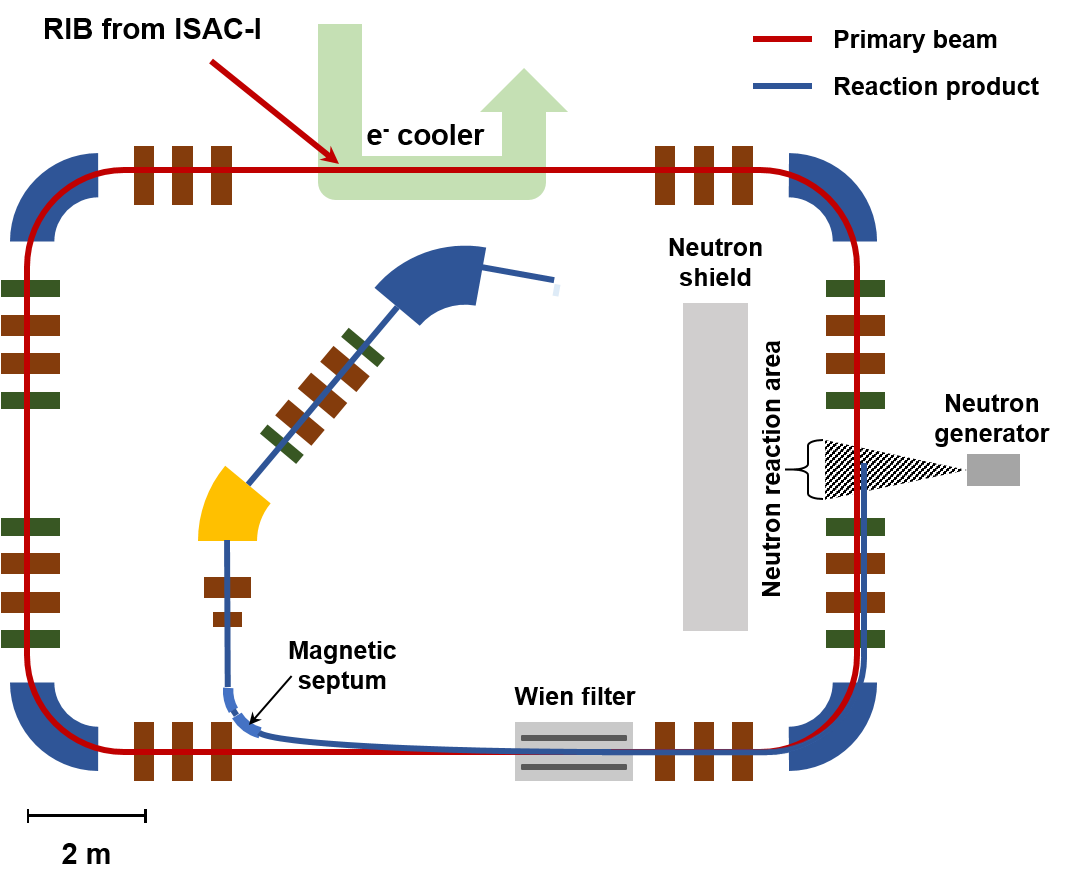}
\caption{Schematic view of the ion storage ring and the recoil separator.
} \label{fig:schematic}
 \end{figure}

To measure neutron-induced radiative capture reactions using inverse kinematics, the idea of an ion storage ring with a reactor as the neutron source was suggested. Based on this idea, a storage ring for TRIUMF, called the TRISR, was proposed and conceptually designed as a combination of an ion storage ring, a velocity filter, a compact neutron generator, and a recoil separator. This system could utilize the existing space and radioactive ion beams available at the ISAC facility of TRIUMF. If the flux is sufficient, the reactor could be replaced with a compact neutron generator. Considering the space limitations of the ISAC-I facility, a storage ring with a Wien filter was designed and verified through ion optical calculations and particle-tracking simulations. The trajectories of the beam and fusion products separated by the velocity filter were calculated and used to determine the position of the magnetic septum used for recoil extraction. The phase space distributions of the recoils at the exit of the extraction system were calculated, and a recoil separator that disperses incident ions according to $m/q$ was conceptually designed. An electrostatic deflector and a dipole magnet are used to separate the incident recoils. Figure \ref{fig:schematic} depicts a schematic of the storage ring, recoil separator, and other essential devices. It is possible that the neutron capture cross-sections of radioactive ions could be measured directly by further developing this conceptual design and building the facility.

\begin{acknowledgments}
This work was supported by the GRA-NRF Program in Canada through the National Research Foundation of Korea (NRF), funded by the MSIT (2021K1A3A1A73090699) and Mitacs Globalink Research Award(IT26791). BD acknowledges the generous support from the Natural Sciences and Engineering Research Council of Canada. TRIUMF receives federal funding via a contribution agreement through the National Research Council of Canada.
\end{acknowledgments}


\clearpage

\bibliography{trisr}

\end{document}